\documentclass[12pt,a4]{article} 
\newcommand{\vs}{\vspace{-0.25cm}}
\usepackage{amsmath,graphicx}
\begin{document}
\hfill JLAB-THY-18-2862
\vspace{-0.5cm}
\begin{center}
 {\Large{\bf Spectral functions of nucleon form factors: \\ Three-pion
continua at low energies}\footnote{This work 
has been supported in part by the DFG excellence cluster ``Origin and Structure
of the Universe'', by DFG and NSFC (CRC110), the U.S. Department of
Energy (contract DE-AC05-06OR23177) and the National Science Foundation
(PHY-1714253).}}  

\medskip

 N. Kaiser$^1$ and E. Passemar$^{2,3,4}$ \\
\end{center}
{\small 1) Physik-Department T39, Technische Universit\"{a}t M\"{u}nchen,
D-85747 Garching, Germany\\ 2) Department of Physics, Indiana University,
Bloomington, IN 47405, USA\\ 3) Center for Exploration
of Energy and Matter, Indiana University, Bloomington, IN 47408, USA \\
4) Theory Center, Thomas Jefferson National Accelerator Facility, Newport News, VA 23606, USA}

\smallskip

\begin{abstract}
We study the imaginary parts of the isoscalar electromagnetic and isovector axial form factors of the nucleon close to the $3\pi$-threshold in covariant baryon chiral perturbation theory. At the two-loop level, the contributions arising from leading and next-to-leading order chiral $\pi N$-vertices, as well as pion-induced excitations of virtual $\Delta(1232)$-isobars, are calculated. It is found that the heavy baryon treatment overestimates substantially these $3\pi$-continua. From a phenomenological analysis, that includes the narrow 
$\omega(783)$-resonance or the broad $a_1$-resonance, one can recognize small windows near threshold, where chiral $3\pi$-dynamics prevails. However, in the case of the isoscalar electromagnetic form factors $G_{E,M}^s(t)$, the radiative correction provided by the 
$\pi^0\gamma$-intermediate state turns out to be of similar size. 
\end{abstract}

\section{Introduction}
The structure of the nucleon as revealed in elastic electron-nucleon and (anti)neutrino-nucleon scattering is encoded in four electromagnetic form factors $G_{E,M}^{p,n}(t)$ and two axial form 
factors $G_{A,P}(t)$, with $t$ the squared momentum-transfer. Dispersion theory is a tool to interpret (and cross check) these scattering data in a largely model independent way \cite{mergell,ina1}. The nucleon form factors are assumed to satisfy unsubtracted dispersion relations and their absorptive parts are often parametrized by a few vector meson poles. However, such an approach is not in conformity with general constraints from unitarity and analyticity. In particular, the singularity structure of the $\pi\pi N$-triangle diagram leads to a pronounced enhancement of the isovector electromagnetic spectral functions on the left wing of the $\rho(770)$-resonance. The two-pion intermediate state can actually be treated exactly (in the energy region $2m_\pi<\sqrt{t}<1\,$GeV) in terms of the pion charge form factor $F_\pi(t)$ and the p-wave $\pi N$ partial wave amplitudes $f^1_\pm(t)$ in the crossed 
$t$-channel $\pi\pi \to \bar N\!N$. For the latter quantities improved results have been obtained in recent dispersion analyses of $\pi N$-scattering \cite{roysteiner,rpuzzle} based on solutions of the Roy-Steiner equations. The calculation of the isovector electromagnetic spectral functions
Im$G_{E,M}^v(t)$ in chiral perturbation theory up to two loops \cite{2loop} is able to account (step by step) for the strong enhancement above the $2\pi$-threshold (which originates from a 
logarithmic singularity at $t_c=4m_\pi^2-m_\pi^4/M^2 = 3.978\,m_\pi^2$ on the second Riemann sheet), but an additional (adjusted) $\rho(770)$-resonance contribution is necessary to reproduce reasonably the empirical spectral functions. On the other hand, the isoscalar electromagnetic form factors $G_{E,M}^s(t)$ are usually represented by sums of a few vector meson poles ($\omega, \phi, s_1, s_2,s_3$) \cite{ina1,ina2}. The effects of the $3\pi$-continuum on the spectral functions Im$G_{E,M}^s(t)$ close to threshold have been calculated in ref.\,\cite{3piHB} using heavy baryon chiral perturbation theory, and it was concluded that these are small against the tails of an $\omega(783)$-resonance with constant width. In that work \cite{3piHB} the axial spectral function Im$G_A(t)$ near the $3\pi$-threshold was also computed and the analogous calculation for the induced pseudoscalar form factor $G_P(t)$ has been performed in 
ref.\,\cite{psinduced}. 

The purpose of the present paper is an improved calculation of the $3\pi$-continua in covariant
baryon chiral perturbation theory. This seems appropriate in view of the expected size of relativistic corrections: $\sqrt{t}/M > 3m_\pi/M=0.44$. In addition to leading order chiral
$\pi N$-vertices, we consider also the next-leading order ones (involving the low-energy constants $c_1, c_2, c_3, c_4$) and we treat the pion-induced excitation of the low-lying $\Delta(1232)$-resonance, described by a Rarita-Schwinger spinor. Our paper is organized as follows. In section 2 we recapitulate the Cutkosky cutting rule applied to two-loop diagrams with a $3\pi$-absorptive part and we present the Lorentz-invariant $3\pi$-phase space integral in explict form together with all kinematical variables. The formulas to project out individual nucleon form factors from the transition matrix elements of the vector and axial-vector currents are also given. Section 3 is devoted to the presentation and discussion of the calculated $3\pi$-spectral function Im$G_{E,M}^s(t)$ and Im$G_A(t)$, separated into contributions from leading order chiral $\pi N$-vertices, next-to-leading order ones, and the inclusion of explicit $\Delta$-isobars. In each case we give also convenient formulas, which refer to the non-relativistic approximation. In section 4 we perform a simple phenomenological analysis by considering the narrow $\omega(783)$-resonance for Im$G_{E,M}^s(t)$ and the broad $a_1(1260)$-resonance for Im$G_A(t)$. In the first case this draws our attention to electromagnetic effects and are thus compelled to compute the radiative correction to Im$G_{E,M}^s(t)$ provided by the $\pi^0\gamma$-intermediate state. The paper ends with a summary and conclusions in section 5.   

\section{Calculation of three-pion spectral functions}
We follow the (standard) definitions for the electromagnetic and axial form factors of the nucleon as given in section 2 of ref.\,\cite{3piHB}, and remind
that each form factor is assumed to satisfy an unsubtracted dispersion relation of the form:
\begin{equation} F(t) = {1\over \pi} 
\int_{t_0}^\infty \!\! dt' \, {\text{Im} F(t') \over t'-t-i\epsilon}\,. \end{equation}
The threshold $t_0$ for hadronic intermediate states is $t_0 =4m_\pi^2$ for the 
isovector electromagnetic form factors $G_{E,M}^v(t)$ and the scalar form factor $\sigma_N(t)$ \cite{2loop}, while  $t_0 =9m_\pi^2$ for the isoscalar electromagnetic form factors $G_{E,M}^s(t)$ and the isovector axial form factors $G_{A,P}(t)$. The measurable electromagnetic form factors of the proton and neutron are composed of the isoscalar and isovector ones as:
\begin{equation}G_{E,M}^{p,n}(t) = G_{E,M}^{s}(t) \pm G_{E,M}^{v}(t)\,,
\end{equation}
where the normalizations $G_E^{s,v}(0)=1/2$, $G_M^s(0)=0.440 $ and $G_M^v(0)=2.353$ hold at $t=0$.

\begin{figure}[ht]\centering
\includegraphics[width=0.4\textwidth]{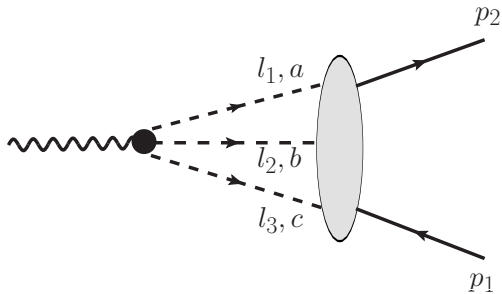}
\caption{Generic two-loop diagram $\gamma^*\to 3\pi\to \bar N\!N$ generating the 
three-pion spectral function.}
\end{figure}   

Figure\,1 shows a generic two-loop diagram with a $3\pi$-intermediate state contributing to the 
nucleon transition matrix element of the vector (or axial-vector) current. The three pions have four-momenta $l_1, l_2, l_3$ and $a, b, c$ are their (cartesian) isospin-indices. Exploiting (perturbative) unitarity in the form of the Cutkosky cutting rules, one obtains for the imaginary (or absorptive) part of the corresponding two-loop amplitude:
\begin{equation} \text{Im}\,{T_{\text{2-loop}} = -{1\over 2} \int\!\!
d\Phi_3 A\!\cdot\!B}\,, \end{equation}
where $A$ denotes the  S-matrix for $\gamma^*\to 3\pi$ and $B$ the S-matrix for $3\pi
\to \bar N\! N$ (in the subthreshold region $\sqrt{t}<2M$). The integral $\int\!d\Phi_3$ goes over the Lorentz-invariant three-pion phase space, whose volume is determined by the kinematical variable 
$t=(l_1+l_2+l_3)^2=(p_2-p_1)^2$ and the pion mass $m_\pi$. In the center-of-mass frame the phase space integration can be expressed as a four-dimensional integral over two energies $(\omega_1, \omega_2)$ and two
angular variables $(x,\varphi)$ by:
\begin{equation}\int\!\!d\Phi_3H(\dots) = {1\over 64\pi^4}
\int\!\!\!\!\int_{z^2<1}\!\!d\omega_1d\omega_2\int_{-1}^1\!\!dx
\int_0^\pi \!\! d\varphi\, H(\dots)\,, \end{equation}
where  $z$ is determined by energy and momentum conservation as
\begin{equation} |\vec l_1||\vec l_2|\,z=\omega_1 \omega_2 -\sqrt{t}
(\omega_1+\omega_2)+{t+m_\pi^2\over 2}\,, \qquad |\vec l_{1,2}|=
\sqrt{\omega_{1,2}^2-m_\pi^2}\,,\end{equation}
and $\sqrt{t}>3m_\pi$ denotes the three-pion invariant mass. The directional cosines 
\begin{equation} x= \hat l_1\!\cdot\! \vec v, \qquad y =\hat l_2\!\cdot\!
  \vec v=xz+\sqrt{(1-x^2)(1-z^2)}\cos\varphi\,, \end{equation}
refer to a unit-vector $\vec v$, which is introduced by the momentum of the
nucleon. The integration region in the $\omega_1\omega_2$-plane, specified by $z^2<1$ in eq.(4), lies
inside a cubic curve and has the explicit boundaries: $\omega_2^- < \omega_2<\omega_2^+$,
with $2\omega_2^\pm = \sqrt{t}-\omega_1\pm |\vec l_1| \big[(t-2\sqrt{t} \omega_1-3m_\pi^2)/(t-2\sqrt{t}
\omega_1+m_\pi^2)\big]^{1/2}$, and $m_\pi<\omega_1<(t-3m_\pi^2)/2\sqrt{t}$.

Before eqs.(3,4) can be applied to calculate spectral functions, one has to project the (individual) electric and magnetic form factors out of the (isoscalar) current transition matrix element
\begin{equation} \bar u_2V^\mu u_1 =  \bar u_2\Big[\gamma^\mu F_1(t)
+{i\over 2M}\sigma^{\mu\nu} (p_2\!-\!p_1)_\nu F_2(t)\Big]u_1\,, \end{equation}
where $\bar u_2$ and $u_1$ are free Dirac-spinors. This is done by multiplying $V^\mu$ with on-shell projectors $\gamma\!\cdot\!p_{2,1}+M$ and taking suitable Dirac-traces:
\begin{eqnarray} && G_E(t) =  F_1(t) + {t\over 4M^2}  F_2(t) =
{(p_1+p_2)_\mu \over 4M(4M^2-t)}{\rm tr}\big\{(\gamma\!\cdot\!p_2+M)V^\mu(\gamma\!\cdot\!p_1+M)\big\}\,,\\  && G_M(t) =  F_1(t) + F_2(t)  =
{1\over 4t}{\rm tr}\Big\{(\gamma\!\cdot\!p_2+M)V^\mu(\gamma\!\cdot\!p_1\!+\!M)\Big[\gamma_\mu +{2M\over t-4M^2}(p_1\!+\!p_2)_\mu\Big]\Big\}\,,
\end{eqnarray}
where $M=939\,$MeV denotes the (average) nucleon mass. The axial and pseudoscalar form factors are projected out of the isovector axial-current transition matrix element (proportional to $\tau_d/2$):
\begin{equation} \bar u_2A^\mu u_1 = \bar u_2\Big[\gamma^\mu G_A(t) +{(p_2 -p_1)^\mu\over 2M}G_P(t)\Big]\gamma_5u_1\,, \end{equation}
in a similar way:
\begin{eqnarray} && 
G_A(t)=-{1\over 4(4M^2-t)} {\rm tr}\Big\{(\gamma\!\cdot\!p_2+M)A^\mu(\gamma\!\cdot\!p_1+M)\Big[ \gamma_\mu +{2M\over t}(p_2-p_1)_\mu\Big]\gamma_5\Big\}\,,
\\ && G_P(t) =  {M\over t(4M^2-t)}{\rm tr}\Big\{(\gamma\!\cdot\!p_2+M)A^\mu(\gamma\!\cdot\!p_1+M)\Big[M \gamma_\mu +\Big({6M^2\over t}-1\Big)(p_2-p_1)_\mu\Big]\gamma_5 \Big\}\,.\end{eqnarray} 
In our calculation the S-matrices $A(\gamma^*\to 3\pi)$ and $B(3\pi\to \bar N\! N)$ are built from chiral vertices and hence the integrand $H(\dots)$ in eq.(4) becomes a rational function of the Lorentz scalar products:
\begin{eqnarray} && l_1\!\cdot\!l_2 = \sqrt{t}(\omega_1+\omega_2)-{t+m_\pi^2
\over 2}\,, \qquad p_1\!\cdot\!p_2 = M^2-{t\over 2}\,,\\ && l_1\!\cdot\!p_{1,2}
= {1\over 2}\Big(\mp \sqrt{t}\,\omega_1 -i x\,\sqrt{4M^2-t}\,|\vec l_1| \Big)\,,
\\ && l_2\!\cdot\!p_{1,2} = {1\over 2}\Big( \mp \sqrt{t}\,\omega_2 -i y\,
\sqrt{4M^2-t}\,|\vec l_2|\Big) \,. \end{eqnarray}
We note that in the nonrelativistic limit $M\to \infty$ the nucleon propagators become complex-valued distributions: 
\begin{equation}{-1\over ix-\epsilon}=\pi\delta(x)+i\text{P}{1\over x}\,, \qquad  {-1\over(i x-\epsilon)(iy+\epsilon)}=\text{P}{1\over x}\text{P}{1\over y}+\pi^2\delta(x) \delta(y)+i\pi \bigg[ \text{P}{1\over x}\delta(y)-\delta(x)\text{P}{1\over y} \bigg]\,, \end{equation}
and the angular integrations $\int_{-1}^1\!dx\int_0^\pi\!d\varphi$ can (and must) be performed analytically. For example, the outcomes of the two distributions in eq.(16) are $\pi^2$ and $2\pi(1-z^2)^{-1/2}\arccos(-z)$.

\section{Results of digrammatic calculation and discussion}

In this section we present the results for the spectral functions Im$G_{E,M}^s(t)$ and Im$G_A(t)$ as calculated from leading-order chiral $\pi N$-vertices, next-to-leading order ones, and pion-induced excitations of virtual $\Delta(1232)$-isobars. Since one works at all three stages with the same couplings of the external sources to three pions, we recapitulate these first. The momentum-dependent (anomalous) coupling of the virtual photon to three out-going pions $\pi^a(l_1), \pi^b(l_2),\pi^c(l_3)$ reads \cite{bijnens}:
\begin{equation} {\epsilon_{abc}\over 4\pi^2 f_\pi^3}
  \epsilon_{\mu\nu\alpha\beta}l_1^\nu l_2^\alpha l_3^\beta\,, \end{equation}
where the charge factor $e$ has been dropped, and $f_\pi=92.2$\,MeV is the pion
decay constant. On the other hand the momentum-dependent coupling of the axial source (with isospin-index $d$) to three (out-going) pions $\pi^a(l_1), \pi^b(l_2),\pi^c(l_3)$  is given by \cite{review}:
\begin{equation}{1\over f_\pi}\big[\delta_{ad}\delta_{bc}(l_2+l_3-l_1)^\mu+
  \delta_{bd}\delta_{ac}(l_1+l_3-l_2)^\mu+\delta_{cd}\delta_{ab}(l_1+l_2-l_3)^\mu
  \big]\,. \end{equation}
\begin{figure}[ht] \begin{center}\includegraphics[width=9cm,clip]{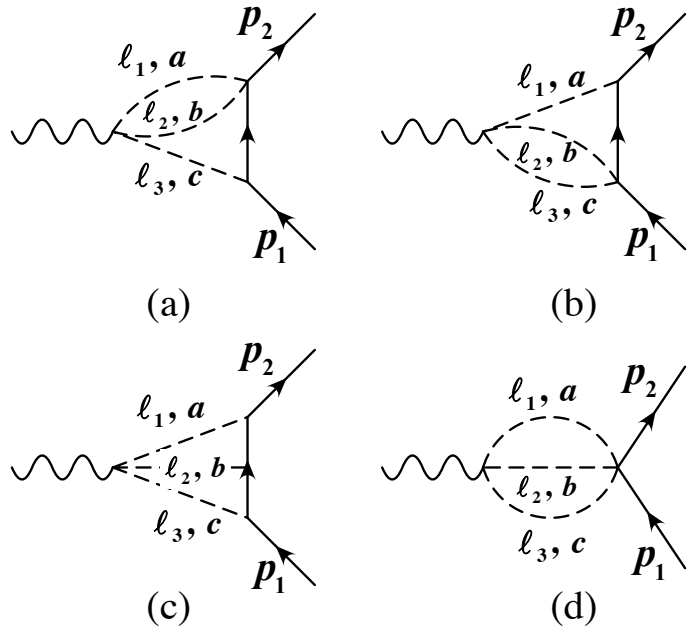}
\end{center}\vspace{-0.6cm}\caption{Two-loop diagrams contributing to the current matrix elements under consideration. Their respective combinatoric factors are $1/2, 1/2, 1,$ and $1/6$.}\end{figure} 
\subsection{Leading-order chiral vertices}

\begin{figure}[ht] \begin{center}\includegraphics[width=9cm,clip]{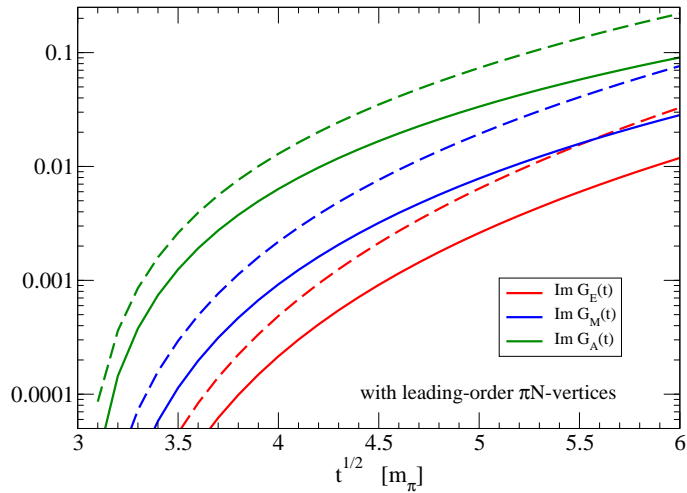}
\end{center}\vspace{-0.6cm}\caption{Spectral functions Im$G_E^s(t)$ (red), Im$G_M^s(t)$ (blue), and Im$G_A(t)$ (green) calculated with leading order chiral $\pi N$-vertices. The dashed lines correspond to the nonrelativistic approximation.}
\end{figure} 
The four relevant two-loop diagrams that need to be evaluated are shown in Fig.\,2. In addition to the well-known pseudovector $\pi N$-coupling and the (vectorial) Weinberg-Tomozawa vertex, one encounters for the axial form factor $G_A(t)$ the chiral contact-vertex with three pions $\pi^a(l_1), \pi^b(l_2), \pi^c(l_3)$ absorbed on a nucleon (see right half of diagram (d) in Fig.\,2): 
\begin{equation} -{g_A \over 4f_\pi^3} \big[ \tau_a \delta_{bc}\, \gamma\!\cdot\!
(l_2+l_3)+ \tau_b \delta_{ac}\, \gamma\!\cdot\!(l_1+l_3)+\tau_c \delta_{ab}\,
\gamma\!\cdot\! (l_1+l_2)\big]\gamma_5  \,. \end{equation}
The rational integrand-functions $H(\dots)$ resulting from the Dirac-traces turn out to be quite lengthy,\footnote{A code with these expressions can be obtained from the authors upon request.} and are therefore not reproduced here. The four-dimensional phase space integration in eq.(4) has been performed numerically, setting $g_A=1.3$ (to have a strong $\pi N$-coupling constant $g_{\pi NN}= g_A M/f_\pi =13.24$) and taking an average pion mass $m_\pi = 138\,$MeV. The obtained leading order spectral functions Im$G_E^s(t)$, Im$G_M^s(t)$ and Im$G_A(t)$ in the low-energy region $3m_\pi <\sqrt{t}<6m_\pi$ are shown in Fig.\,3 by red, blue and green lines, respectively. A logarithmic scale is used to make visible the very small values in the threshold region. The dashed lines in Fig.\,3 correspond to the nonrelativistic approximation. For the leading terms in the $1/M$-expansion of the electromagnetic spectral functions one can actually give convenient formulas:  
\begin{eqnarray} \text{Im}G_E^s(t)&=& {g_A\over 8M(8\pi)^4f_\pi^6 \sqrt{t}}
\int_{2m_\pi}^{\sqrt{t}-m_\pi} \!\!dw\,(w^2-4m_\pi^2)^{3/2} \lambda(w,t) \nonumber
\\ &&+{3g_A^3t\over(4\pi)^5f_\pi^6}\int\!\!\!\! \int_{z^2<1}\!\!d\omega_1d\omega_2
     \,|\vec l_1||\vec l_2|\sqrt{1-z^2}\arccos(-z)\,,\end{eqnarray}
\begin{equation} \text{Im} G_M^s(t)={g_AM \over 4(8\pi)^4f_\pi^6 t^{3/2}}
\int_{2m_\pi}^{\sqrt{t}-m_\pi} \!\!dw\,\sqrt{w^2-4m_\pi^2}\Big[w^2-4m_\pi^2+g_A^2
(5w^2-8m_\pi^2)\Big] \lambda(w,t)\,,\end{equation}
with the abbreviation $\lambda(w,t)=[t-(w+m_\pi)^2][t-(w-m_\pi)^2]$ and $w$ denotes a $2\pi$-invariant mass. The prefactor $M$ in Im$G_M^s(t)$ originates from the magnetic coupling $i\vec \sigma\!\times\!(\vec p_2-\vec p_1)/2M$. The analogous (nonrelativistic) formula for Im$G_A(t)$ can be found in eq.(26) of ref.\,\cite{3piHB} and that for Im$G_P(t)$ in eqs.(4,5) of ref.\,\cite{psinduced}. We do not discuss here  Im$G_P(t)$, since this form factor is dominated (at low momentum transfers) by the pion-pole term $G_P(t)^{(\pi)} = 4g_{\pi NN} M f_\pi/(m_\pi^2-t)$. Note that eq.(20) includes in the first line also the leading $1/M$-correction to Im$G_E^s(t)$ coming from the two diagrams (a) and (b) in Fig.\,2 proportional to $g_A$. By inspection of Fig.\,3 one observes that the heavy baryon treatment (used in ref.\,\cite{3piHB}) leads to an overestimation of the spectral functions near the $3\pi$-threshold by about a factor of 2 to 3. This noteworthy property of the chiral $3\pi$-continua points to sizeable relativistic corrections of magnitude $\sqrt{t}/M > 3m_\pi/M=0.44$.    

\subsection{Second-order chiral vertices}
Next, we compute the spectral functions with vertices from the second-order chiral 
$\pi N$-Lagrangian. The pertinent S-matrix for the absorption of two pions $\pi^a(l_1), \pi^b(l_2)$ on a nucleon reads: 
\begin{equation} -{2i\over f_\pi^2}\delta_{ab}\Big[ 2c_1 m_\pi^2+{c_2\over 4M^2}
(p+p')\!\cdot\!l_1(p+p')\!\cdot\!l_2+c_3 \,l_1\!\cdot\!l_2\Big]+ {i c_4\over
f_\pi^2}\epsilon_{abe}\tau_e \sigma_{\mu\nu}l_1^\mu l_2^\nu\,, \end{equation}
where $p$ and $p'$ denote in-going and out-going nucleon four-momenta. This $2\pi$-contact vertex enters now diagrams (a) and (b) in Fig.\,2. Note that due to the contraction with $\epsilon_{abc}$ only the (last) $c_4$-term contributes to Im\,$G_{E,M}^s(t)$, whereas all four $c_i$-terms contribute to Im\,$G_{A,P}(t)$. For the numerical evaluation of $\int\!d\Phi_3\,H(\dots)$, we choose first the (rounded) values  $c_1=-0.8$\,GeV$^{-1}$, $c_2=3.3$\,GeV$^{-1}$, $c_3=-4.7$\,GeV$^{-1}$ and $c_4 = 3.4$\,GeV$^{-1}$  of the second-order low-energy constants \cite{buett}. Similar values are often employed in N$^3$LO chiral NN-potentials and they are consistent with recent determinations from $\pi N$-dispersion relation analyses \cite{cideterm} or fits of chiral $\pi N$-amplitudes to pion-nucleon scattering phase shifts \cite{oller}. With this chosen input the results for the spectral functions Im$G_E^s(t)$, Im$G_M^s(t)$ and Im$G_A(t)$ are shown in Fig.\,4. One sees that these (formally) subleading corrections are roughly of similar size as the leading order terms displayed in Fig.\,3. A more detailed comparison reveals that the $c_4$-contribution to Im$G_E^s(t)$ is suppressed for $3m_\pi<\sqrt{t}<5m_\pi$ and this suppression is more  pronounced for Im$G_M^s(t)$. On the other hand the combined $c_i$-contributions to Im$G_A(t)$ exceed the leading order axial spectral function already for $\sqrt{t} >3.7 m_\pi$. The latter feature is explained by the large value of the low-energy constant $c_3$. The dashed lines in Fig.\,4 refer to the nonrelativistic approximation, which again leads to an overestimation by about a factor 2. In the nonrelativistic limit  the following integral-representations can be derived for the electromagnetic spectral functions:
\begin{equation} \text{Im}G_E^s(t)= {g_A c_4\over 2(8\pi)^4f_\pi^6 \sqrt{t}}
\int_{2m_\pi}^{\sqrt{t}-m_\pi}\!\!dw\,(w^2-4m_\pi^2)^{3/2}\lambda(w,t)\,,
\end{equation}
\begin{equation} \text{Im}G_M^s(t)= {g_A c_4 M\over 2(4\pi)^5 f_\pi^6 t}
\int_{2m_\pi}^{\sqrt{t}-m_\pi} \!\!dw\,(w^2-4m_\pi^2)^{3/2} \sqrt{\lambda(w,t)}
\Big[t- w^2-m_\pi^2- {\lambda(w,t)\over 3t}\Big]\,, \end{equation}
and for the axial spectral functions:
\begin{eqnarray}
\text{Im}G_A(t) & = & {g_A \over (4f_\pi)^4 \pi^2 \sqrt{t}} 
\int_{2m_\pi}^{\sqrt{t} -m_\pi}\!\!dw\,\sqrt{w^2-4m_\pi^2}\, \Big\{{2c_4\over 3}
(w^2-4m_\pi^2) (t-w^2-m_\pi^2) \nonumber \\ && +{\lambda(w,t) \over t} \Big[
c_3(2m_\pi^2-w^2)-4c_1m_\pi^2 +{c_2+c_4\over 6}(4m_\pi^2-w^2)\Big] \Big\}\,,
\end{eqnarray}
\begin{eqnarray} \text{Im}G_P(t) & = & {g_A M^2 \over 
64\pi^2 f_\pi^4 (t-m_\pi^2)t^{3/2}} \int_{2m_\pi}^{\sqrt{t} -m_\pi}\!\!dw\,
\sqrt{w^2-4m_\pi^2}\, \Big\{\Big[ c_3(2m_\pi^2-w^2)-4c_1 m_\pi^2\nonumber \\ && 
+{c_2+c_4\over 6}(4m_\pi^2-w^2)\Big] \Big[{3m_\pi^2 \over t}(w^2-m_\pi^2)^2
-3w^2 m_\pi^2 -m_\pi^4-\lambda(w,t)\Big] \nonumber \\ &&+{c_4\over 6}(w^2-
 4m_\pi^2)\big[t(4w^2+5m_\pi^2)-4t^2-w^2m_\pi^2+m_\pi^4\big]\Big\}\,,
\end{eqnarray}
where the latter expression includes also pion-pole diagrams (axial source\,$\to\pi\to 3\pi\to \bar N\!N$) involving the chiral $\pi\pi$-interaction. As a good check, one can verify that the combination Im$G_A(t) +(t/ 4M^2)$Im$G_P(t)$, related to the divergence of the isovector axial-current, scales as $m_\pi^2$. Note that the $dw$-integrals in eqs.(20,21,23,25,26) can be solved in terms of square-root and logarithmic functions.
\begin{figure}[ht] \begin{center}\includegraphics[width=9cm,clip]{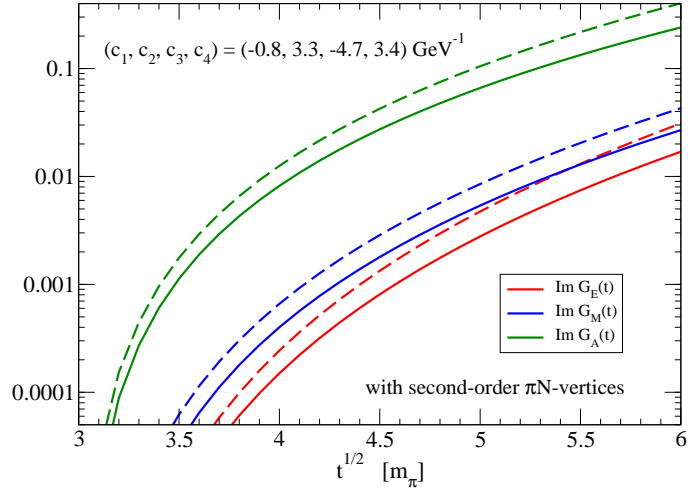}
\end{center}\vspace{-0.6cm}\caption{Spectral functions Im$G_E^s(t)$, Im$G_M^s(t)$, and Im$G_A(t)$ calculated with second-order chiral $\pi N$-vertices for low-energy constants $(c_1, c_2, c_3, c_4)= (-0.8, 3.3, -4.7, 3.4)\,$GeV$^{-1}$. The dashed lines correspond to the nonrelativistic approximation.}
\end{figure}
\begin{figure}[ht] \begin{center}\includegraphics[width=9cm,clip]{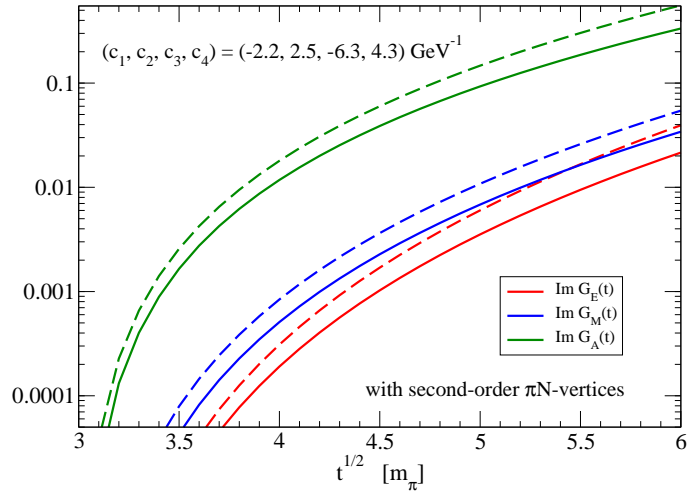}
  \end{center}\vspace{-0.6cm}\caption{Spectral functions Im$G_E^s(t)$, Im$G_M^s(t)$, and Im$G_A(t)$ calculated with second-order chiral $\pi N$-vertices for low-energy constants $(c_1, c_2, c_3, c_4)= (-2.2, 2.5, -6.3, 4.3)\,$GeV$^{-1}$. The dashed lines correspond to the nonrelativistic approximation.}\end{figure}

The previously used low-energy constants $c_i$ stem from determinations at the one-loop level of chiral
perturbation theory. Therefore, we employ as an alternative the values $(c_1, c_2, c_3, c_4)= (-2.2, 2.5,
-6.3, 4.3)\,$GeV$^{-1}$, which were deduced in ref.\cite{cialternative} from a covariant tree-level
calculation of $\pi N$-scattering, including constraints from the inelastic processes $\pi N\to
\pi\pi N$. The corresponding results for the spectral functions Im$G_E^s(t)$, Im$G_M^s(t)$ and Im$G_A(t)$
are shown in Fig.\,5. By comparison to Fig.\,4, one recognizes for the electromagnetic spectral functions
Im$G_{E,M}^s(t)\sim c_4$ the obvious enhancement factor 1.26 from the larger $c_4$-value, while the axial
spectral function Im$G_M^s(t)$ increased by roughly a factor 1.5. Besides this weak enhancement the
pattern of curves in Fig.\,4 and Fig.\,5 is the same. At this point one should also note that
$c_1=-2.2\,$GeV$^{-1}$ gives (at tree-level) a nucleon sigma-term of $\sigma_N = -4c_1 m_\pi^2=167\,$MeV,
which exceeds the empirical value by about a factor 3.
\subsection{Inclusion of explicit $\Delta(1232)$-isobars}
The sizeable magnitude of the low-energy constants $c_{2,3,4}$ is explained by large contributions from the $\Delta(1232)$-resonance, which strongly couples to the $\pi N$-system. The covariant description of the $\Delta(1232)$-isobar with spin and isospin $3/2$ requires a Rarita-Schwinger spinor field $\Psi_\alpha$. In this formulation the spin-$3/2$ propagator (vector-index $\beta$ to $\alpha$) takes the (common) form \cite{review}: 
\begin{equation} {i\over 3} {\gamma\!\cdot\!P+M_\Delta \over M_\Delta^2-P^2}
\bigg(3g_{\alpha\beta}-\gamma_\alpha \gamma_\beta-{2P_\alpha P_\beta \over M_\Delta^2} +{P_\alpha \gamma_\beta\!-\!\gamma_\alpha P_\beta\over M_\Delta} \bigg)\,, \end{equation}
with $P$ the four-momentum of the propagating $\Delta(1232)$-isobar. In order to keed the two-loop calculations tractable, we choose minimal forms of the vertices for the coupling of an in-going pion to $\Delta N$ and $\Delta\Delta$, which read:
\begin{equation}\Delta_\alpha N\pi^a(l_1)\!: \quad -{3g_A\over 2\sqrt{2}f_\pi}l_1^\alpha
T_a\,, \qquad\qquad \Delta_\alpha\Delta_\beta\pi^b(l_2)\!: \quad {3g_A\over 10f_\pi}
g^{\alpha\beta} \gamma\!\cdot\!l_2 \gamma_5\,\Theta_b\,.\end{equation}
The isospin transition operator $T_a$ satisfies the relation $T_aT_b^\dagger = (2\delta_{ab}-i\epsilon_{abc}\tau_c)/3$, and for the isospin-3/2 operator  
$\Theta_a$ (a $4\times 4$ matrix) the reduction formula $T_a\Theta_bT_c^\dagger = (5i\epsilon_{abc}-\delta_{ab} \tau_c+4\delta_{ac}\tau_b-\delta_{bc}\tau_a)/3$ is relevant. The coupling constants in eq.(28) obey the ratios $g_{\pi N\Delta}/g_{\pi NN} = 3/\sqrt{2}$ and $g_{\pi \Delta\Delta}/g_{\pi NN} = 1/5$ as inferred from large-$N_c$ QCD \cite{deltadelta}. One should note that extended versions of the vertices in eq.(28) with further off-shell parameters have been proposed \cite{review,deltadelta}, but these parameters are not well determined. Since no direct empirical information is available, the relation $g_{\pi \Delta\Delta}=
g_{\pi NN}/5$ is commonly used \cite{deltadelta}. Alternative and more sophisticated approaches to treat
the $\Delta(1232)$-isobar in chiral perturbation theory (e.g. small-scale expansion and $\delta$-counting) have been developed in refs.\cite{hemmert,pasc1,pasc2,pasc3}.

\begin{figure} \begin{center}\includegraphics[width=9cm,clip]{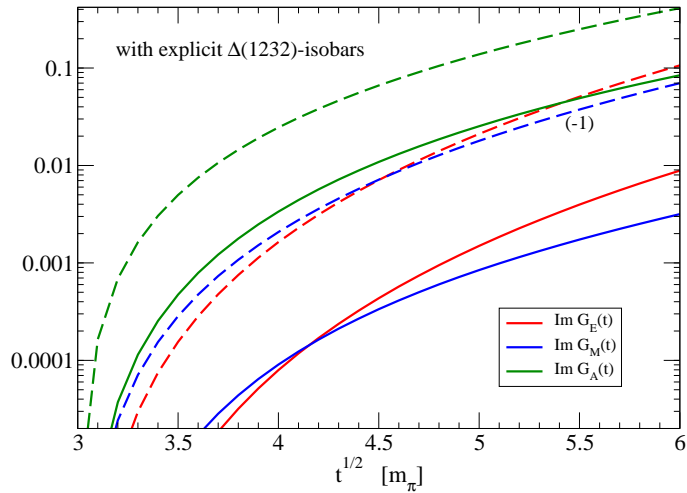}
\end{center}\vspace{-0.6cm}\caption{Spectral functions Im$G_E^s(t)$, Im$G_M^s(t)$, and Im$G_A(t)$ calculated from diagrams with single and double propagation of 
$\Delta$-isobars. The dashed lines refer to the nonrelativistic approximation.}
\end{figure}

Employing the just described formulation of vertices and propagators, we have derived the (extremely
lengthy) integrand-functions $H(\dots)$ for Im$G_{E,M}^s(t)$  and Im$G_A(t)$ from the diagrams with single and double virtual $\Delta(1232)$-excitation (analogs of diagram (c) in Fig.\,2).
The corresponding numerical results are shown by the full lines in Fig.\,6.
By comparison with Fig.\,4 one observes that the spectral functions get appreciably reduced by the energy-dependent $\Delta$-propagators. The suppression factor is about  2 to 3 for  Im$G_E(t)$ and Im$G_A(t)$, whereas it amounts to about 7 to 8 for Im$G_M(t)$. Of course, the physics here and in subsection 3.2 is somewhat different. The $c_i$-parameters represent more than the $\Delta$-intermediate state ($c_2^{(\Delta)} = -c_3^{(\Delta)} = 2 c_4^{(\Delta)} \simeq 2.9\,$GeV$^{-1}$) and there are partly compensating effects from single and double $\Delta$-isobar excitation. It is also instructive to present formulas which refer to the nonrelativistic approximation. For doing that we take first the limit of $\Delta N$-degeneracy, $M_\Delta=M $, and then expand in $1/M$. This way one obtains for the electric spectral function:
\begin{equation} \text{Im}G_E^s(t) ={3g_A^3 t\over (4\pi)^5 f_\pi^6} \Big(1+
{5\over2} \Big) \int\!\!\!\!\int_{z^2<1}\!\!d\omega_1d\omega_2\,|\vec l_1|
|\vec l_2|\sqrt{1\!-\!z^2}\arccos(-z)\,, \end{equation}
where the factor $(1+5/2)$ displays the separate contributions from $\Delta 
N$ and $\Delta\Delta$. Likewise, one finds for the magnetic spectral function:
\begin{equation} \text{Im} G_M^s(t)={g_A^3M \over 4(8\pi)^4f_\pi^6 t^{3/2}}
\int_{2m_\pi}^{\sqrt{t}-m_\pi} \!\!dw\,\sqrt{w^2-4m_\pi^2}(8m_\pi^2-5w^2)
\lambda(w,t)\,,\end{equation}
which is opposite to the term proportional to $g_A^3$ in eq.(21). This opposite sign and the factor $(1+5/2)$ can be deduced from the spin- and isospin-algebra involved in the (nonrelativistic) three-pion to nucleon coupling, which has to be spin-independent (spin-dependent) for the electric (magnetic) form
factor. The dashed lines in Fig.\,6 correspond to the nonrelativistic approximations written in eqs.(29,30) as well as to a more complicated formula for 
$\lim_{M_\Delta=M\to\infty}\text{Im}G_A(t)$. One can see that the proposed nonrelativistic approximation strongly overestimates the results for spectral functions with 
$\Delta(1232)$-excitation based on fully relativistic kinematics. In the case of Im$G_M^s(t)$ there is even a difference in sign.

\section{Phenomenological analysis and $\pi^0\gamma$ intermediate state}
In this section we want to find out the low-energy region, where the $3\pi$-continua calculated in covariant baryon chiral perturbation theory could become physically relevant. For that purpose we compare our results with the spectral functions produced by the respective lowest-lying vector-meson resonance. For the isoscalar electromagnetic form factors $G_{E,M}^s(t)$ this is obviously the narrow $\omega$-meson with mass $m_\omega=783\,$MeV and decay width $\Gamma_\omega=8.5\,$MeV\,$=(7.6+0.7+0.2)\,$MeV \cite{pdg}. In this decomposition of  $\Gamma_\omega$ the first two entries refer to the dominant decay modes $\omega\to \pi^+\pi^0\pi^-$ and  $\omega\to \pi^0\gamma$. The reasonable assumption of 
$\omega(783)$-meson dominance in the region $3m_\pi<\sqrt{t}<7m_\pi$ leads to the following complex-valued form factors: 
\begin{equation} G_{E,M}^s(t)={(0.50,0.44)m_\omega^2 \over m_\omega^2-t-i 
 m_\omega\Gamma_\omega(t)}\,, \end{equation}
with $\Gamma_\omega(t)$ an energy-dependent $\omega$-meson decay width. The
numbers $0.50$ and $0.44$ in the numerator of eq.(31) are the isoscalar charge
and isoscalar magnetic moment of the nucleon. Modelling the two dominant decay
modes by appropriate contact-couplings, one gets:
\begin{equation}\Gamma_\omega(t)={h^2\over m_\omega t}\int_{2m_\pi}^{\sqrt{t}-m_\pi}\!\!dw \big[(w^2-4m_\pi^2)\lambda(w,t)\big]^{3/2}+ {h'^2\over m_\omega t}(t-m_\pi^2)^3 \,, \end{equation}
with the parameters $h = 2.72$\,GeV$^{-3}$ and $h' = 0.040$\,GeV$^{-1}$ adjusted to the partial
decay widths. We note as an aside that with this modelling of $\Gamma_\omega(t)$ the denominator
in eq.(31) becomes zero at $t= (32.17-0.340\, i)m_\pi^2$,
corresponding to a complex $\omega$-meson pole at $\sqrt{t} = (782.8-4.14\, i)\,$MeV.
\begin{figure} \begin{center}\includegraphics[width=9cm,clip]{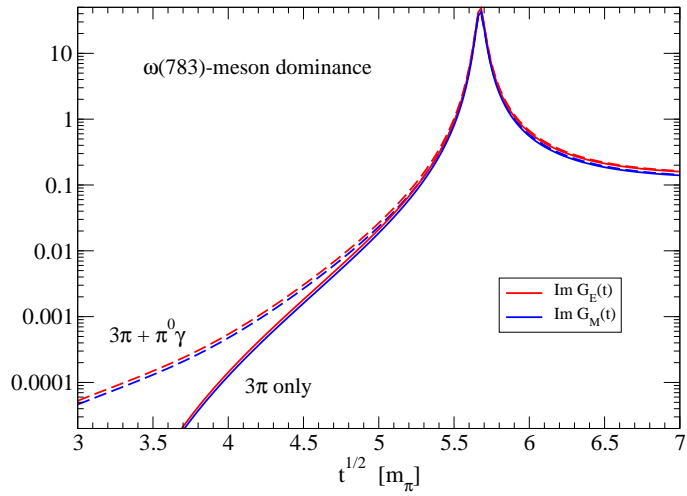}
\end{center}\vspace{-0.6cm}\caption{Isoscalar electromagnetic spectral functions 
Im$G_{E,M}^s(t)$ assuming  $\omega(783)$-meson dominance.}
\end{figure}

The resulting imaginary parts Im$G_{E,M}^s(t)$ are shown in Fig.\,7. The resonance curves for the electric and magnetic form factor are almost equal, due to similar normalizations $0.50 \simeq 0.44$. One sees that in the region $3m_\pi < \sqrt{t}<5m_\pi$ the $3\pi$-only contributions from the $\omega(783)$-resonance fall below the (combined) chiral $3\pi$-continua, whereas the additional $\pi^0\gamma$-mode introduces appreciable strength in the threshold region.  In view of this striking effect, one is compelled to compute the radiative correction to the isoscalar electromagnetic spectral functions coming from the 
$\pi^0\gamma$-intermediate state. The pertinent S-matrix for $\pi^0\to \gamma\gamma$ reads: $(-i \alpha_\text{em} /\pi f_\pi) \epsilon_{\mu\nu\alpha\beta}
k_1^\alpha k_2^\beta$, where $(k_1,\mu)$ and $(k_2,\nu)$ pertain to out-going photons. The one-loop calculation of both diagrams $\gamma^*\to\pi^0\gamma\to\bar 
N\!N$ requires only one angular integration $(m_\pi^2-t)/(32\pi t)\int_{-1}^1\!
dx$, such that the $\pi^0\gamma$-contribution to the isoscalar electromagnetic 
spectral functions can be given in analytical form:
\begin{equation}\text{Im}G_E^s(t)={\alpha_\text{em} g_A (t-m_\pi^2)^2 \over
(4\pi f_\pi)^2(4M^2-t)} \bigg\{ -{1\over 4} +{\kappa_v \over 3}\Big({t-m_\pi^2
\over 16M^2}-1 +{m_\pi^2\over 4t}\Big)+{(1+ \kappa_v)M^2 \over \sqrt{t(4M^2-t)}}
\arccos{ \sqrt{t}\over 2M}\bigg\}\,, \end{equation}
\begin{equation}\text{Im}G_M^s(t) = {\alpha_\text{em} g_A M^2(t-m_\pi^2)^2 \over
(4\pi f_\pi)^2 (4M^2-t)t} \bigg\{{1\over 2} +{\kappa_v\over 3}\Big(2-{t+2m_\pi^2 
\over 8M^2}+{m_\pi^2 \over t}\Big) +{4M^2-(2+\kappa_v)t\over 2\sqrt{t(4M^2-t)}} 
\arccos{\sqrt{t}\over 2M}\bigg\}\,, \end{equation}
with $\kappa_v = \kappa_p-\kappa_n=3.706$ the (large) isovector anomalous magnetic moment and $\alpha_\text{em}=1/137$. Note that one averages here over proton and neutron form factors, while the $\pi^0N$-coupling $\sim \tau_3$ introduces an opposite sign for the magnetic moment term.  The curves resulting from the expressions in 
eqs.(33,34), with threshold behavior Im$G_{E,M}^s(t)\sim (t-m_\pi^2)^2$, are drawn in Fig.\,8. One observes that in the region 
$3m_\pi<\sqrt{t}<4m_\pi$ the radiative corrections due to the $\pi^0\gamma$-intermediate state exceed the (leading order) chiral $3\pi$-continua (dashed-dotted lines in Fig.\,8). This behavior is explained kinematically by the fast decrease of the $3\pi$-phase space towards the threshold $\sqrt{t}=3m_\pi$, while the $\pi^0\gamma$-phase space remains open down to 
$\sqrt{t}=m_\pi$.  

\begin{figure}[ht]
\begin{center}\includegraphics[width=9cm,clip]{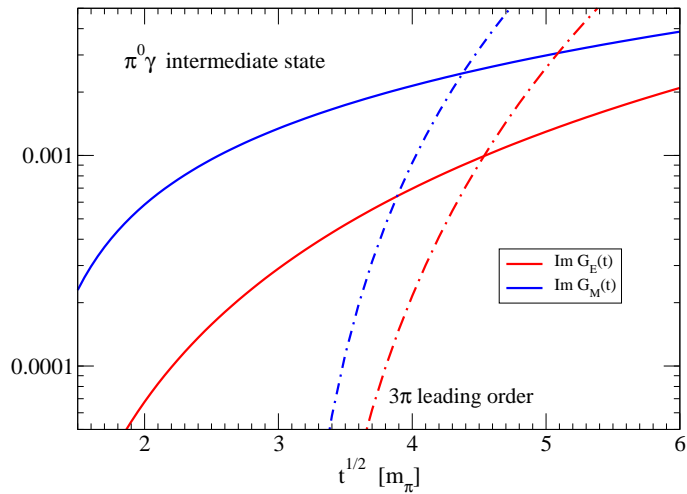}
\end{center}\vspace{-0.6cm}\caption{Contributions to the isoscalar electromagnetic spectral functions Im$G_{E,M}^s(t)$ from the $\pi^0\gamma$-intermediate state compared to leading order $3\pi$-spectra.}\end{figure}

\begin{figure}
\begin{center}\includegraphics[width=9cm,clip]{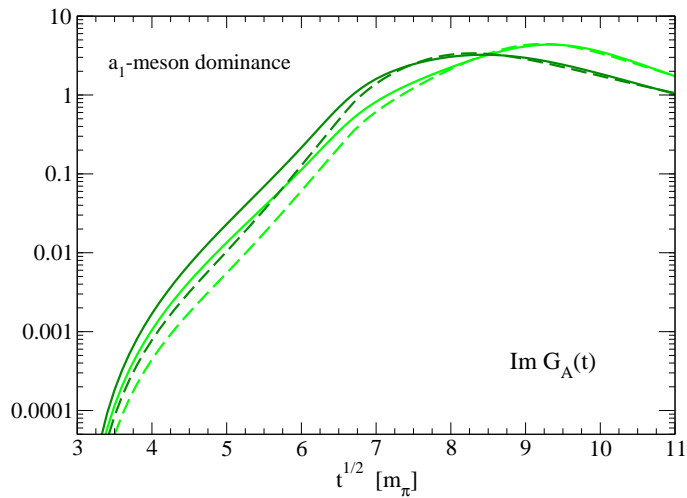}
\end{center}\vspace{-0.6cm}
\caption{Axial spectral function Im$G_A(t)$ assuming $a_1$-meson dominance. The two sets of (light and dark green) curves correspond to masses and widths of
$(m_{a_1},\Gamma_{a_1})=(1.3, 0.38)\,$GeV \cite{compass} and  $(m_{a_1},\Gamma_{a_1})=(1.2,0.48)\,$GeV \cite{dumm1}.}
\end{figure}

In analogy to eq.(31) the nucleon axial form factor $G_A(t)$ dominated by the $a_1$-resonance reads: 
\begin{equation}G_A(t)={g_A m_{a_1}^2\over m_{a_1}^2-t-im_{a_1}\Gamma_{a_1}(t)}\,,\end{equation}
with the (proper) axial-vector coupling constant $g_A=1.27$ \cite{pdg}. The mass and width of the broad $a_1$-meson are still under debate, due to conflicting results from different experiments. A very recent partial wave analysis of diffractive dissociation data ($\pi^- p\to \pi^-\pi^+\pi^- p$)  by the COMPASS collaboration \cite{compass} finds the (central) values $m_{a_1} = 1.3$\,GeV and  $\Gamma_{a_1} =0.38$\,GeV. On the other hand the values extracted from  $\tau$-lepton decays in ref.\cite{dumm1} are  $m_{a_1} = 1.2$\,GeV and  $\Gamma_{a_1} = 0.48$\,GeV, while a later reanalysis in ref.\,\cite{dumm2} gave a somewhat lower $a_1$-mass of $m_{a_1} = 1.12$\,GeV. Moreover, the Joint Physics Analysis Center
Collaboration \cite{jpacc} extracted from the ALEPH data on $\tau^-\to \pi^-
\pi^+\pi^-\nu_\tau$ a  complex $a_1$-pole position of $m_{a_1}-i \Gamma_{a_1}/2 =(1.21-0.29\,i)\,$GeV. The  model employed in ref.\cite{jpacc} is based on approximate three-body unitary and the singularity structures related to $\pi\pi$-subchannel resonances were carefully addressed.

The full lines in Fig.\,9 show the axial spectral function Im$G_A(t)$ using the
specific form of $\Gamma_{a_1}(t)$, which follows from integrating (interfering) Breit-Wigner functions for the $\rho(770)$-resonance over the $3\pi$-phase
space (see section 3 in ref.\cite{dumm2}).  The dashed lines were obtained with the phenomenological parametrization of $\Gamma_{a_1}(t)$ from ref.\,\cite{kuehn}, which describes separately the regions below and above the $\rho\pi$-threshold $t=(m_\rho+m_\pi)^2$.  The light and dark pair of curves refer to the parameter sets $(m_{a_1},\Gamma_{a_1}) =(1.3, 0.38)\,$GeV \cite{compass} and $(m_{a_1},\Gamma_{a_1}) =(1.2, 0.48)\,$GeV \cite{dumm1}, which are clearly distinguished by their shifted peaks. By comparison with the full (green) lines in Figs.\,3 and 4 one can recognize an energy window near threshold, $3m_\pi<\sqrt{t}<5m_\pi$, in which the chiral $3\pi$-continua do prevail. However, such tiny contributions to the axial spectral function are presumably irrelevant for physical observables.    

\section{Summary and conclusions}
In this work we have studied the imaginary parts of the isoscalar electromagnetic and isovector axial form factors of the nucleon close to the $3\pi$-threshold. The contributions to Im$G_{E,M}^s(t)$ and Im$G_A(t)$ arising from chiral $\pi N$-vertices at leading and next-to-leading order, as well as pion-induced $\Delta(1232)$-excitations have been calculated and compared with each other. It was found that the heavy baryon approach overestimates these chiral $3\pi$-continua substantially. Moreover, leading and next-to-leading order contributions
to the chiral $3\pi$-continua are of similar size, due to the large low-energy constants $c_{1,2,3,4}$.  From a phenomenological analysis, that included the narrow $\omega(783)$-resonance or the broad $a_1(1260)$-resonance, one could recognize small windows near threshold, where chiral $3\pi$-dynamics prevails. However, for Im$G_{E,M}^s(t)$ the radiative correction provided by the $\pi^0\gamma$-intermediate state becomes actually more relevant in the region close to threshold. Although the net result of our covariant calculation of the $3\pi$-spectral functions in chiral perturbation theory is still uncertain, one can nevertheless conclude that these chiral $3\pi$-continua for the nucleon form factors are too weak to influence physical observables in a significant way.  

\end{document}